\documentclass[a4paper,11pt]{article}
\usepackage{jheppub} 

\usepackage{jheppub} 
\usepackage{mathrsfs,amscd}
 \usepackage{bm,bbm}
 \usepackage{graphicx}
 \usepackage{float}
 \usepackage{xcolor}
 \usepackage{caption, subcaption}
 
 \usepackage{tikz}

\def\Z{\mathbb{Z}}

\def\beq#1\eeq{\begin{align}#1\end{align}}

\title{Hyperelliptic families and 4d $\mathcal{N}=2$ SCFT }

\author[a]{Dan Xie}
\author[b]{, Zekai Yu}

\affiliation[a]{Department of Mathematics, Tsinghua University, Beijing, 100084, China}
\affiliation[b]{Qiuzhen College, Tsinghua University, Beijing, 100084, China}

\abstract{We classify four dimensional $\mathcal{N}=2$ SCFTs whose Seiberg-Witten (SW) geometries can be written as hyperelliptic families. 
By using special Kähler condition of SW geometry, we reduce the problem to one parameter quasi-homogeneous hyperelliptic families $y^2=f(x,t)$. The classification is given by further demanding that the complex algebraic surface defined by $y^2=f(x,t)$  has an isolated singularity. We then write down the full SW geometry by looking at mini-versal deformations of the one parameter family, and the SW differential is also written down. The detailed physical data for these theories are found by matching the theory with other known construction. Our solutions recover the known rank one and rank two results, and 
give some infinite sequences valid at arbitrary ranks. We also studied $Z_2$ quotient of above hyperelliptic families which give rise to $B$ type and $D$ type conformal gauge theory, and further generalizations.}

\begin{document} 
\maketitle
\flushbottom

\section{Introduction}
The Coulomb branch solution of a general $\mathcal{N}=2$ theory can be described by 
finding a family of algebraic varieties $F(z_i, u_i, \lambda_i, m_i)=0$  fibered over the space of generalized Coulomb branch parameters
involving expectation value $u_i$ of Coulomb branch operators, mass parameters $m_i$,
and coupling constants $\lambda_i$ (including relevant couplings and exactly marginal deformations). 
One also need to find a SW differential $\lambda_{sw}$ to describe the central charge of BPS particles.
The combination of varieties $F$ and $\lambda_{sw}$ are called SW geometry, see \cite{Seiberg:1994rs,Seiberg:1994aj} for the
SW curves \footnote{The SW geometries for some 4d $\mathcal{N}=2$ theories must be described by a high dimensional algebraic varieties.} of $SU(2)$ gauge theory.

The  SW geometry is usually found through string/M theory-motivated methods \cite{Witten:1997sc,gaiotton2} and 
the connection of SW solution to integrable system \cite{DonagiWitten,Martinec:1995by}.
Most SW geometries were discovered by following two methods:
a) The spectral curve of an integral system, and the most well studied ones are the so-called Hitchin system \cite{Hitchin:1987mz,DonagiWitten,Nanopoulos:2009uw,Xie:2012hs, Wang:2015mra,Wang:2018gvb};
b) The mini-versal deformation of an isolated three dimensional canonical singularities \cite{xie20154d,Wang:2016yha}.

While a large number of SW geometries can be found using the methods above, 
one should not forget that the resulting geometries are often written in a complicated way which makes 
it hard to perform a detailed study of the low energy physics. This is in particular true for 
some SW geometries engineered using Hitchin system. Often an equivalent but simpler SW geometry would make 
the physical exploration much more viable. 

Holding this in mind, here we try to classify SW geometry which can be expressed in terms of the simplest algebraic curves, i.e. the hyperelliptic curves:
\begin{equation*}
y^2=f(x, u_i, \lambda_i, m_i);
\end{equation*}
Here $f$ is a degree $2g+1$ or $2g+2$ polynomial in $x$, and $u_i, i=1,\ldots, g$ denotes the expectation value of 
Coulomb branch operators. We focus on the SW geometry for the SCFT, and perform the classification through the following steps:
\begin{enumerate}
\item In the first step, we turn off the mass parameters and coupling constants, so that the geometry $y^2=f(x,u_i)$ have 
a $\mathbb{C}^*$ action to reflect the $U(1)_R$ symmetry of a SCFT. 
\item We then impose the special Kähler condition which implies that the derivative of SW differential $\lambda_{sw}$
should be in the space of holormorphic differential of the hyperelliptic curve \cite{Xie:2021hxd}.
\begin{equation*}
{\partial \lambda_{sw} \over \partial u_i}\in H^{1,0}.
\end{equation*}
The simplest solution would indicate that the SW curve and SW differential are written in the form 
\ref{swcurve} and \ref{swdifferential}.
\item The above two considerations reduce the classification into finding out possible one-parameter quasi-homogeneous families $y^2=f(x,t)$.
We impose a further condition, namely the surface singularity defined by the equation is isolated, and so 
a complete classification of one-parameter geometries is possible, see \ref{oneparameter} for the resulting 8 types of geometries.
\item We then find the full SW geometry by looking at the versal deformations of one parameter geometry, and then replace the parameter $t$ by
a polynomial in $x$ which involves VEV of Coulomb branch operators and possibly other coupling constants.
\end{enumerate}
The detailed studies of those theories are listed in table. [\ref{typeI},\ref{typeII},\ref{typeIII},\ref{typeIV},\ref{typeV},\ref{typeVI},\ref{typeVII},\ref{typeVIII},\ref{type0}]. 
Our list recovers the rank one \cite{Argyres:1995xn,Minahan:1996cj} and rank two results \cite{Argyres:20221}, and generalizes them to arbitrary rank.

Two further generalizations are studied: first we study the $\Z_2$ quotient of the hyperelliptic families and 
get three classes of theories, see Table. \ref{twist1} and \ref{twist2}. These can be thought of as the arbitrary rank-generalization
of the rank one $I_4$ series studied in \cite{Argyres:2015gha}, and include the $B$ type conformal gauge theory.
We then also study the case where the square root of the holomorphic differential is used, which would describe 
$D$ type conformal gauge theory, and generalizations.

This paper is organized as follows: section 2 gave a detailed illustration of our classification strategy 
and also the detailed physical properties of those theories, by matching them with known constructions; 
$\Z_2$ quotients of the curves are studied in section 3; section 4 discusses 
the SW differential and the $D$ type gauge theory; A conclusion is given in section 5.

\section{Hyperelliptic families and SCFTs}
First we assume that the SW curve of a 4d $\mathcal{N}=2$ SCFT is given by a family of genus $g$ hyperelliptic curves:
\begin{equation}
   y^2=f(x,u_i)
\end{equation}
where $f(x,u_i)$ is a polynomial in $x$ of degree $2g+1$ or $2g+2$. 
Here $u_i,~i=1,\ldots, g$ denote the Coulomb branch parameters. To get the SW geometry of a SCFT, the above curve needs to have a $\mathbb{C}^*$ action 
acting on $(x,y,u_i)$ which is the incarnation of the (complexified) $U(1)_R$ symmetry of the SCFT.

The SW differential is given by  $\lambda_{sw}=R(x,y,u_i) dx$ with $R$ a rational function of $(x,y,u_i)$, and the derivative of $\lambda_{sw}$ with respect to Coulomb branch operator is \cite{brieskorn2012plane}:
\begin{equation}
   {\partial \over \partial u_i}\int_\lambda \lambda_{sw} = {\partial \over \partial u_i} \int_\lambda R dx=\int ({\partial R \over \partial u_i}-{\partial R \over \partial y}({\partial F \over \partial y})^{-1} {\partial F\over \partial u_i}) dx;
\end{equation}
Here $\lambda$ is the cycle on Riemann surface, and $F$ is the equation $F=y^2-f(x,u_i)=0$. The special Kähler condition requires that \cite{Xie:2021hxd}:
\begin{equation}
{\partial \lambda_{sw} \over \partial u_i} \subset H^{1,0}(C);
\end{equation}
Here $H^{1,0}(C)$ denotes the space of holomorphic differential of the hyperelliptic curve $C$, which has a canonical basis 
\begin{equation*}
\omega_i=\frac{x^i dx}{y},~~i=0,\ldots, g-1.
\end{equation*}

Let $t$ denote the Coulomb branch operator with maximal scaling dimension, and assume that 
the derivative of SW differential with respect of $t$ is proportional to $\frac{dx}{y}$, i.e.
\begin{equation}
    {\partial \lambda_{sw} \over \partial t}=({\partial R \over \partial t}-{\partial R \over \partial y}({\partial F \over \partial y})^{-1} {\partial F\over \partial t}) dx =c {dx \over y};
    \label{basic}
\end{equation}
Here $c$ is a nonzero constant. Now if the SW curve takes the following special form 
\begin{equation}
   \boxed{ y^2=f(x, u_{g-1}x^{g-1}+u_{g-2}x^{g-2}+ \ldots+ u_1x+t).}
   \label{swcurve}
\end{equation}
and similarly the SW differential takes the form 
\begin{equation}
    \boxed{\lambda_{sw}=R(x,y, u_{g-1}x^{g-1}+u_{g-2}x^{g-2}+ \ldots+ u_1x+t) dx.}
    \label{swdifferential}
\end{equation}
Then we have 
\begin{equation}
  {\partial \lambda_{sw} \over \partial u_i}=({\partial R \over \partial u_i}-{\partial R \over \partial y}({\partial F \over \partial y})^{-1} {\partial F\over \partial u_i}) dx= x^i {\partial \lambda_{sw} \over \partial t}=c {x^i dx \over y};
  \label{deriva}
\end{equation}
Here we used the fact ${\partial R\over \partial u_i}=x^{i} {\partial R\over \partial t}$, and ${\partial F\over \partial u_i}=x^{i} {\partial F\over \partial t}$. So the special Kähler condition is satisfied automatically for the SW curve $\ref{swcurve}$ and SW differential $\ref{swdifferential}$, 
as long as the equation \ref{basic} has a solution (we will discuss this issue in section 4). 

\textbf{One parameter curve}:  By looking at the curve \ref{swcurve} and \ref{swdifferential}, one can see that the SW family is determined 
by the curve with just one parameter, which corresponds to the Coulomb branch operator of maximal scaling dimension. 
The classification of theories (or SW geometries) is 
now reduced to the classification of the one parameter hyperelliptic curve $y^2=f(x,t)$ with a $\mathbb{C}^*$ action. We impose a further constraint on the one parameter family:  There is an isolated singularity (of the complex surface defined by $y^2=f(x,t)$) at the origin $x=y=t=0$. 
Such one parameter hyperelliptic families can be easily classified as follows:
\begin{align}
&I: y^2=x^{2g+1}+t^a,~~\nonumber\\
&II: y^2=x^{2g+2}+t^a,~~\nonumber\\
&III: y^2=x(x^{2g}+t^a),~~\nonumber\\
&IV: y^2=x(x^{2g+1}+t^a),~~\nonumber\\
& V: y^2=t(x^{g+2}+t^a),~~\nonumber\\
& VI: y^2=t(x^{g+3}+t^a),~~\nonumber\\
& VII: y^2=xt(x^{g+1}+t^a),~~\nonumber\\
& VIII: y^2=xt(x^{g+2}+t^a).~~
\label{oneparameter}
\end{align}
\textbf{Proof}: The isolated quasi-homogeneous singularity of two variables $(x,t)$ is classified as $x^a+t^b, x^a+xt^b,~~x^a t+y^b t$ \cite{arnold1998singularity}, so one has the above 8 types.

\textbf{Coulomb branch spectrum}: Now one can proceed to the calculation of Coulomb branch spectrum of the theory. The basic idea is that the SW differential $\lambda_{sw}$ gives the mass of BPS particle and so it has scaling dimension one. The equation \ref{deriva} gives a relation for the scaling dimension
\begin{equation*}
1-[u_i]=i [x]-[y];
\end{equation*}
We find that 
\begin{equation*}
[u_{i-1}]-[u_i]=[x];
\end{equation*}
So if the scaling dimension of $x$ is positive, the maximal scaling dimension of Coulomb branch operators is $[t]$ and the gap of the Coulomb branch spectrum is given by $[x]$. 
The $\mathbb{C}^*$ action of the curve $y^2=f(x,t)$ yields two equations relating the scaling dimensions, which would help us to completely determine the spectrum.

\textit{Example}: Let's consider  one parameter curve $y^2=x^{2g+1}+t^2$. We have the following three equations
\begin{align*}
& 2[y]=(2g+1)[x], \nonumber\\
& 2[y]=2[t], \nonumber\\
&1-[t]=[x]-[y].
\end{align*}
and so we find $[t]=\frac{g+1}{2},~~[x]=1$, so the Coulomb branch spectrum is $\frac{3}{2}, \frac{5}{2}, \ldots, \frac{2g+1}{2}$.

\textbf{Full SW curve}: One can turn on exactly marginal deformations, relevant deformations, and 
mass deformations without lifting the Coulomb branch. These are encoded in the deformations of above SW curves. They can be derived as follows:
\begin{enumerate}
\item To get the SW curve with general Coulomb branch deformations, we replace the parameter $t$ in the one parameter curve by a degree $g-1$ polynomial in $x$. More generally, substitution by a polynomial in higher degree in $x$ is possible, as long as the scaling dimensions of the parameters that come with the higher order terms are non-negative:
\begin{equation*}
t \to  \ldots+ \lambda_2 x^{g+1}+\lambda_1 x^g+ u_{g-1}x^{g-1}+\ldots+t.
\end{equation*}
The parameters $\lambda_1, \lambda_2,\ldots$ includes mass parameters, relevant deformations, and exactly marginal deformations. Extra care must be taken to ensure that the above replacement do not affect the genus of the curve. 
\item Mass deformations can be found by finding the mini-versal deformation of the one parameter curve. The mini-versal deformation 
of an isolated singularity of our type is defined as follows
\begin{equation}
y^2=f(x,t)+\sum_\alpha \lambda_\alpha \phi_\alpha;
\end{equation}
Here $\phi_\alpha$ is the monomial basis of the Jacobi algebra defined as follows
\begin{equation*}
J_f={C[x,t] \over \{ {\partial f\over \partial x}, {\partial f \over \partial t}\}}.
\end{equation*}
The scaling dimension of $\lambda_\alpha$ could be larger than or equal to one. 
Parameters with scaling dimension integral and larger than one should be interpreted as the Casimiers of the non-abelian flavor symmetry.

\end{enumerate}

\textit{Example}: Let's consider the one parameter curve $y^2=x(x^{2g}+t^2)$, and the scaling dimensions are $[x]=2, [t]=2g$.
The basis of Jacobi algebra $J_f$ is
\begin{equation*}
J_f=\{t, x^{2g},x^{2g-1},\ldots, 1\};
\end{equation*}
and so there are a total of $2g+2$ mass parameters.
The deformed one parameter curve is 
\begin{equation*}
y^2=x^{2g+1}+xt^2+[T_2 x^{2g}+T_4 x^{2g-1}+\ldots+T_{4g+2}]+T_{2g+2}^{'} t.
\end{equation*}

Now the replacement of $t$ (with maximal number of deformations) is
\begin{equation*}
t\to \tau x^{g}+u_{g-1} x^{g-1}+\ldots+ t;
\end{equation*}
We see that there is an exactly marginal deformation represented by . 
The full SW curve is 
\begin{align*}
&y^2=x^{2g+1}+x(\tau x^g+u_g x^{g-1}+\ldots+t)^2 +[T_2 x^{2g}+T_4 x^{2g-1}+\ldots+T_{4g+2}] \nonumber\\
&+T_{2g+2}^{'}(\tau x^{g}+u_{g-1} x^{g-1}+\ldots+ t).
\end{align*}
and the scaling dimension of the mass deformations $(T_2, T_4, \ldots, T_{4g+2}, T_{2g+2}^{'})$ is $(2,4,\ldots, 4g+2, 2g+2)$,
so the flavor symmetry is $SO(4g+4)$.

\subsection{Detailed data of SCFTs: type I-IV }
Now we study those SCFTs whose SW curve is given by the form \ref{swcurve}. The one parameter family \ref{oneparameter} is $I, II, III, IV$.
These families share the property that the one parameter family already has genus $g$.

We list the Coulomb branch spectrum,  the central charge, flavor symmetry, and the number of $A_1$ singularities for the generic deformations. 
The corresponding periodic map \cite{xie2023pseudoperiodic}. Factorization of the related mapping class group element is also listed (see \cite{ishizakapresent}), which would be important to understand the low energy physics \cite{Xie:2023fact}. The $(A_1, G)$ sequence was first found in \cite{Eguchi:1996vu}, and the class $S$ construction was first found in \cite{gaiotton2}.

\begin{table}[H]
\caption{Type I:~~$y^2=x^{2g+1}+t^a$.}
\begin{center}
\begin{tabular}{|c|c|c|c|} \hline
Theory &  $(A_1, A_{2g})$ & $D_2(SU(2g+1)$ & ~ \\ \hline
curve & $y^2=x^{2g+1}+t$ & $y^2=x^{2g+1}+t^2$ & $y^2=x^{2g+1}+t^a, a<\frac{4g+2}{2g-1}$ \\ \hline
 CB spectrum &  $\frac{2g+4}{2g+3},\frac{2g+6}{2g+3},\ldots,\frac{4g+2}{2g+3}$ & $\frac{3}{2},\frac{5}{2},\ldots, \frac{2g+1}{2}$ & $a\leq 5,~~g=1$  \\ \hline
 periodic map & $\frac{1}{2(2g+1)}+\frac{1}{2}+\frac{g}{2g+1}$ & $\frac{g}{2g+1}+\frac{g}{2g+1}+\frac{1}{2g+1}$ & $a\leq 3,~~g=2$   \\ \hline
 factorization & $\phi_1=\tau_1 \tau_2 \ldots \tau_{2g}$ & $\phi_1^2$ & $a\leq 2,~~g\geq 3$   \\ \hline
 central charges & $a=\frac{g(24g+19)}{24(2g+3)},c=\frac{g(6g+5)}{6(2g+3)}$ & $a=\frac{7(g^2+g)}{24},c=\frac{g^2+g}{3}$ & ~  \\ \hline
 flavor symmetry & None & $su(2g+1)$ & ~  \\ \hline
 \# $A_1$ singularity & $2g$ & $4g$ & ~  \\ \hline
 \end{tabular}
 \begin{tabular}{|l|l|} \hline
 Full SW curve $a=1$ & $y^2=x^{2g+1}+(\lambda_1 x^{2g-1}+\lambda_2 x^{2g-2}+\ldots+ u_{g-1}x^{g-1}+\ldots+t)$ \\ \hline
  Full SW curve $a=2$ &  $y^2=x^{2g+1}+(T_2 x^{2g-1}+\ldots+ T_{2g+1})+(\lambda_{\frac{1}{2}} x^g+ u_{g-1}x^{g-1}+\ldots+t)^2$\\ \hline 
  Full SW curve $a=3, g=2$ &  $y^2=x^{5}+(u_{4}x+t)^3+(T_{12}x^{3}+T_{18}x^2+T_{24} x+T_{30})$ \\
  flavor $e_8$& $+ (T_2x^3+T_8 x^2+T_{14} x+T_{20})(u_{4}x+t)$\\ \hline 
\end{tabular}
\end{center}
\label{typeI}
\end{table}%

 \begin{table}[H]
\caption{Type II: $y^2=x^{2g+2}+t^a$.}
\begin{center}
\begin{tabular}{|c|c|c|c|c|} \hline
Theory & $(A_1, A_{2g+1})$ & $SU(g+1)-(2g+2)$ & ~\\ \hline
curve & $y^2=x^{2g+2}+t$ & $y^2=x^{2g+2}+ t^2$   & $y^2=x^{2g+2}+t^a$ \\ \hline 
CB spectrum & $\frac{g+3}{g+2}, \ldots, \frac{2g+2}{g+2}$ &  $2,3,\ldots, g+1$ &  $a<\frac{2g+2}{g}$\\ \hline
periodic map & $\frac{1}{2g+2}+\frac{1}{2g+2}+\frac{g}{g+1}$ &  $\frac{1}{g+1}+\frac{1}{g+1}+\frac{g}{g+1}+\frac{g}{g+1}$ &  $a\leq 3, g=1$ \\ \hline
factorization & $\phi_2=\tau_1 \tau_2 \ldots \tau_{2g} \tau_{2g}$  & $\phi_2^2$ & $a\leq 2, g\geq 2$ \\ \hline
central charges & $a=\frac{12g^2+19g+2}{24(g+2)},c=\frac{3g^2+5g+1}{6(g+2)}$ & $a=\frac{7g^2+14g+2}{24},c=\frac{2g^2+4g+1}{6}$ &~ \\ \hline
 flavor symmetry & $u(1)$ & $su(2g+2)\times u(1)$  &~  \\ \hline
 \# $A_1$ singularity & $2g+1$ & $4g+2$ &~ \\ \hline
\end{tabular}
\begin{tabular}{|l|l|} \hline
 Full SW curve $a=1$ & $y^2=x^{2g+2}+(\lambda_1 x^{2g}+\lambda_2 x^{2g-1}+\ldots+ m x^g+u_{g-1}x^{g-1}+\ldots+t)$ \\ \hline
  Full SW curve $a=2$ &  $y^2=x^{2g+2}+(T_2 x^{2g}+ \ldots+ T_{2g+2})+(\tau x^{g+1}+m x^g+ u_{g-1}x^{g-1}+\ldots+t)^2$\\ \hline 
\end{tabular}
\end{center}
\label{typeII}
\end{table}%

 \begin{table}[H]
\caption{Type III: $y^2=x(x^{2g}+t^a)$.}
\begin{center}
\begin{tabular}{|c|c|c|c|c|} \hline
Theory & $(A_1, D_{2g+1})$  & $USp(2g)-(2g+2)$ & ~ \\ \hline
curve & $y^2=x(x^{2g}+t)$ & $y^2=x(x^{2g}+ t^2)$ & $y^2=x(x^{2g}+t^a),~~a<\frac{4g}{2g-1}$ \\ \hline
 CB spectrum & $\frac{2 g+2}{2 g+1}, \ldots,\frac{4 g}{2 g+1}$ & $2,4,\ldots, 2g$ & $a\leq 3, g=1$  \\ \hline
periodic map &  $\frac{1}{4g}+\frac{2g-1}{4g}+\frac{1}{2}$  & $\frac{1}{2}+\frac{1}{2}+\frac{1}{2g}+\frac{2g-1}{2g}$ & $a\leq 2, g\geq 2$   \\ \hline
factorization & $\phi_3=\tau_1 \tau_2 \ldots \tau_{2g} \tau_{2g+1}$  & $\phi_3^2$ & $\phi_3^a$ \\ \hline
 central charges & $a=\frac{g(8g+3)}{8(2g+1)},c=\frac{g}{2}$ & $a=\frac{14g^2+9g}{24},c=\frac{4g^2+3g}{6}$ & ~  \\ \hline
 flavor symmetry & $su(2)$ & $so(4g+4)$ & ~  \\ \hline
 \# $A_1$ singularity & $2g+1$ & $4g+2$ & ~  \\ \hline
\end{tabular}
\begin{tabular}{|l|l|} \hline
 Full SW curve $a=1$ & $y^2=x^{2g+1}+x(\lambda_1 x^{2g-1}+\ldots+ u_{g-1}x^{g-1}+\ldots+t)+T_2$ \\ \hline
  Full SW curve $a=2$ &  $y^2=x^{2g+1}+x(\tau x^g+u_{g-1} x^{g-1}+\ldots+t)^2 +[T_2 x^{2g}+T_4 x^{2g-1}+\ldots+T_{4g+2}]$\\
  ~& $+T_{2g+2}^{'}(\tau x^{g}+u_{g-1} x^{g-1}+\ldots+ t)$  \\ \hline
\end{tabular}
\end{center}
\label{typeIII}
\end{table}%

 \begin{table}[H]
\caption{Type IV: $y^2=x(x^{2g+1}+t^a)$.}
\begin{center}
\begin{tabular}{|c|c|c|c|c|} \hline
Theory & $(A_1, D_{2g+2})$ & Class S & ~ \\ \hline
curve  & $y^2=x(x^{2g+1}+t)$ &$y^2=x(x^{2g+1}+t^2)$  & $y^2=x(x^{2g+1}+t^a)$,~~$a<\frac{2g+1}{g}$  \\ \hline
CB spectrum  & $\frac{g+2}{ g+1},\ldots,\frac{2g+1}{ g+1}$ & $3,5,\ldots, 2g+1$ & $a\leq 2$ \\ \hline
periodic map & $\frac{1}{2 g+1}+\frac{1}{2g+1}+\frac{2 g-1}{2 g+1}$ & $\frac{2g}{2 g+1}+\frac{g+1}{2g+1}+\frac{g+1}{2 g+1}$ & ~ \\ \hline 
factorization & $\phi_4$ & $\phi_4^2$ & $\phi_4^a$ \\ \hline
central charges & $a=\frac{g}{2}+\frac{1}{12},c=\frac{g}{2}+\frac{1}{6}$ & $a=\frac{14g^2+23g+4}{24},c=\frac{4g^2+7g+2}{6}$ & ~  \\ \hline
 flavor symmetry & $su(2)\times u(1)$ & $so(4g+6)\times u(1) (E_6$ if $g=1)$ & ~  \\ \hline
 \# $A_1$ singularity & $2g+2$ & $4g+4$ & ~  \\ \hline
\end{tabular}
 \begin{tabular}{|l|l|} \hline
 Full SW curve $a=1$ & $y^2=x^{2g+2}+x(\lambda_1 x^{2g}+\ldots+ m x^g+ u_{g-1}x^{g-1}+\ldots+t)+T_2$ \\ \hline
  Full SW curve $a=2$ &  $y^2=x^{2g+2}+x(m x^g+u_{g-1} x^{g-1}+\ldots+t)^2 +[T_2 x^{2g+1}+T_4 x^{2g-1}+\ldots+T_{4g+4}]$\\
  ~& $+T_{2g+3}^{'}(m x^{g}+u_{g-1} x^{g-1}+\ldots+ t)$  \\ \hline
\end{tabular}
\end{center}
\label{typeIV}
\end{table}%

\subsection{Detailed data of SCFTs: type V-VIII}
The one parameter families of those theories have genus less than $g$, so that substituting  $t$ by a degree $g-1$ polynomial in $x$ would give us a genus $g$ hyperelliptic family. The periodic map is for 
the one parameter curve.

 \begin{table}[H]
\caption{Type V: $y^2=t(x^{g+2}+t^a)$.}
\begin{center}
\begin{tabular}{|c|c|c|} \hline
Theory & ~ & ~\\ \hline
curve & $y^2=t(x^{g+2}+t)$ &$y^2=t(x^{g+2}+t^a)$,~~$a<\frac{g+2}{g}$ \\ \hline
CB spectrum &  $3,4,\ldots, g+2$ & $a\leq 1$ \\ \hline
Class S& $([g,1^2],[1^{g+2}],[1^{g+2}])$ & $~$ \\ \hline
Periodic map& $\frac{g+1}{g+2}+\frac{g+1}{g+2}+\frac{2}{g+2}$~$g~odd$ &\\ 
~& $\frac{g+1}{g+2}+\frac{g+1}{g+2}+\frac{1}{(g+2)/2}$~$g~even$&\\ \hline
central charges& $a=\frac{7g^2+28g+6}{24},c=\frac{2g^2+8g+3}{6}$ & $~$ \\ \hline
flavor& $su(2g+4)\times su(2) (E_6$ if $g=1)$  & $~$ \\ \hline
\# $A_1$ singularities& $4g+4$ & $~$ \\ \hline
\end{tabular}
 \begin{tabular}{|l|l|} \hline
  Full SW curve $a=1$ &  $y^2=(T_2^\prime x^g+u_{g-1} x^{g-1}+\ldots+t) [x^{g+2}+(T_2^\prime x^g+u_{g-1} x^{g-1}+\ldots+t)]$ \\ 
  & $+[T_2 x^{2g+2}+T_3 x^{2g+1}+\ldots+T_{2g+4}]$\\ \hline
\end{tabular}
\end{center}
\label{typeV}
\end{table}%
 
  \begin{table}[H]
\caption{Type VI: $y^2=t(x^{g+3}+t^a)$.}
\begin{center}
\begin{tabular}{|c|c|c|} \hline
Theory & ~ & ~ \\ \hline
curve & $y^2=t(x^{g+3}+t)$ & $y^2=t(x^{g+3}+t^a)$,~~$a<\frac{g+3}{g+1}$  \\ \hline
CB spectrum & $4,5,\ldots, g+3$ & $a\leq 1$ \\ \hline
Class S&$[(g+1),2], [1^{g+3}],[1^{g+3}]$ & $~$ \\ \hline
Periodic map& $\frac{g+2}{g+3}+\frac{g+2}{g+3}+\frac{2}{g+3}$~$g~even$ &\\ 
& $\frac{g+2}{g+3}+\frac{g+2}{g+3}+\frac{1}{(g+3)/2}$~$g~odd$&\\ \hline
central charges& $a=\frac{7g^2+42g+10}{24},c=\frac{2g^2+12g+5}{6}$ & $~$ \\ \hline
flavor & $su(2g+6)$ & $~$ \\ \hline
\# $A_1$ singularities& $4g+5$ & $~$ \\ \hline
\end{tabular}
 \begin{tabular}{|l|l|} \hline
  Full SW curve $a=1$ &  $y^2=(u_{g-1} x^{g-1}+\ldots+t) [x^{g+3}+(u_{g-1} x^{g-1}+\ldots+t)]$ \\ 
  & $+[T_2 x^{2g+4}+T_3 x^{2g-1}+\ldots+T_{2g+6}]$\\ \hline
\end{tabular}
\end{center}
\label{typeVI}
\end{table}%

  \begin{table}[H]
\caption{Type VII: $y^2=xt(x^{g+1}+t^a)$.}
\begin{center}
\begin{tabular}{|c|c|c|} \hline
Theory & ~ & ~\\ \hline
curve & $y^2=xt(x^{g+1}+t)$ & $y^2=xt(x^{g+1}+t^a)$,~~$a<\frac{g+1}{g}$ \\ \hline
CB spectrum & $4,6,\ldots, 2g+2$ & $a\leq 1$ \\ \hline
Class S&$[(g+1)^2],[g^2,1^2],[1^{2g+2}]$ & $~$ \\ \hline
Periodic map& $\frac{2g+1}{2g+2}+\frac{g+2}{2g+2}+\frac{1}{2}$~$g~odd$ & \\ 
& $\frac{2g+1}{2g+2}+\frac{(g+2)/2}{g+1}+\frac{1}{2}$~$g~even$&\\ \hline
central charges& $a=\frac{14g^2+37g+8}{24},c=\frac{4g^2+11g+4}{6}$ & $~$ \\ \hline
flavor & $so(4g+8)\times su(2)$ & $~$ \\ \hline
\# $A_1$ singularities& $4g+5$ & $~$ \\ \hline
\end{tabular}
 \begin{tabular}{|l|l|} \hline
  Full SW curve $a=1$ &  $y^2=x(T_2^\prime x^g+u_{g-1} x^{g-1}+\ldots+t) [x^{g+1}+(T_2^\prime x^g+u_{g-1} x^{g-1}+\ldots+t)]$ \\ 
  & $+[T_2 x^{2g+2}+T_4 x^{2g+1}+\ldots+T_{4g+6}]+T_{2g+4}^{'}(T_2^\prime x^g+u_{g-1} x^{g-1}+\ldots+t)$\\ \hline
\end{tabular}
\end{center}
\label{typeVII}
\end{table}%

  \begin{table}[H]
\caption{Type VIII: $y^2=xt(x^{g+2}+t^a)$.}
\begin{center}
\begin{tabular}{|c|c|c|} \hline
Theory & ~ & ~\\ \hline
curve & $y^2=xt(x^{g+2}+t)$ & $y^2=xt(x^{g+2}+t^a)$,~~$a<\frac{g+2}{g+1}$ \\ \hline
CB spectrum & $6,8,\ldots, 2g+4$ & $a\leq 1$ \\ \hline
Class S& $[(g+2)^2],[(g+1)^2,2],[1^{2g+4}]$ & $~$ \\ \hline
Periodic map& $\frac{2g+3}{2g+4}+\frac{g+3}{2g+4}+\frac{1}{2}$~$g~even$&\\ 
& $\frac{2g+3}{2g+4}+\frac{(g+3)/2}{g+2}+\frac{1}{2}$~$g~odd$&\\ \hline
central charges& $a=\frac{14g^2+65g+16}{24},c=\frac{4g^2+19g+8}{6}$ & $~$ \\ \hline
flavor & $so(4g+12)$ & $~$ \\ \hline
\# $A_1$ singularities& $4g+6$ & $~$ \\ \hline
\end{tabular}
 \begin{tabular}{|l|l|} \hline
  Full SW curve $a=1$ &  $y^2=x(u_{g-1} x^{g-1}+\ldots+t) [x^{g+2}+(u_{g-1} x^{g-1}+\ldots+t)]+$ \\ 
  & $[T_2 x^{2g+4}+T_4 x^{2g+3}+\ldots+T_{4g+10}]+T_{2g+10}^{'}(u_{g-1} x^{g-1}+\ldots+t)$\\ \hline
\end{tabular}
\end{center}
\label{typeVIII}
\end{table}%

  \begin{table}[H]
\caption{The case when $[x]=0$.}
\begin{center}
\begin{tabular}{|c|c|c|} \hline
Theory & $2-\underbrace{SU(2)-\ldots-SU(2)}_{g}-2$ & ~\\ \hline
curve & $y^2=t(x^{g+3}+\tau_1 x^{g+2}+\ldots+\tau_{g+2})$ & ~ \\ \hline
CB spectrum & $2,\ldots, 2$ & $~$ \\ \hline
Class S& $[1^2]\times (g+3)$ & $~$ \\ \hline
central charges& $a=\frac{19g+4}{24},c=\frac{5g+2}{6}$ & $~$ \\ \hline
Periodic map& $(\frac{1}{2})^{g+4}$& \\ \hline
flavor & $su(2)^{g+3}$ & $~$ \\ \hline
\# $A_1$ singularities& $4g+2$ & $~$ \\ \hline
\end{tabular}
\begin{tabular}{|l|l|} \hline
  Full SW curve &  $y^2=(u_{g-1} x^{g-1}+\ldots+t) [x^{g+3}+\tau_1 x^{g+2}+\ldots+\tau_{g+2})]$ \\ \hline
\end{tabular}
\end{center}
\label{type0}
\end{table}%

\section{$\Z_2$ quotient}
The Coulomb branch solution is given by the fibration of abelian varieties over the generalized 
Coulomb branch \cite{DonagiWitten}. In above studies, the abelian varieties are given by the 
Jacobian of a hyperellliptic curves. Some abelian varities are given by the finite quotient 
of the Jacobian variety, and are called Prym variety. Given above hyperelliptic families, 
we would like to study the finite quotient to get the SW geometries for other SCFTs.

Let's consider the curves which admit a $\Z_2$ action $x\to -x$, then the anti-invariant holomorphic differentials are
\begin{equation*}
{x^{2i} dx \over y},~i=0,1\ldots,
\end{equation*}
So the number would be $\frac{g}{2}$ for even $g$, and $\frac{g+1}{2}$ for $g$ odd. 
Now the special Kähler condition is 
\begin{equation*}
{\partial \lambda_{sw}\over \partial t}=c \frac{ dx}{y};
\end{equation*}
So $\lambda_{sw}$ is required to be odd under $x\to -x$ if we'd like to preserve the $t$ in the spectrum. Once we keep $t$ 
in the spectrum,  the operator $u_i$ with $i$ even would be preserved. The Dirac quantization would put further 
constraint on the $t$ transformation of the curve (we would like to discuss this subtle issue elsewhere).

\textit{Example}: Let's now consider the curve defined as $y^2=x^{2g+2}+t^2$, with $g$ odd. The full SW curve would be 
\begin{equation*}
y^2=x^{2g+2}+(\tau x^{g+1}+u_{g-1} x^{g-1}+u_{g-3}x^{g-3}+\ldots+t)^2+(T_2 x^{2g}+T_4x^{2g-2}+\ldots+T_{2g}x^2).
\end{equation*}
Terms of odd powers of $x$ are eliminated in the last bracket since the curve is required to be invariant under $x\rightarrow -x$. This curve should describe the conformal gauge theory $SO(g+2)$ coupled with $2g$ half-hypermultiplets, so the flavor symmetry is $usp(2g)$. 
Notice that the constant deformation term is not allowed (see \cite{Danielsson:1995is}).

We search all such theories and list them in table. \ref{twist1} and \ref{twist2}. Notice the rank one theories are 
the $I_4$ series found in \cite{Argyres:2015gha}.

 \begin{table}[H]
\caption{$\Z_2$ quotient: $y^2=x^{2g+2}+t^2$.}
\begin{center}
\begin{tabular}{|c|c|c|} \hline
Theory & $SO(g+2)-2g(half-hyper)$ & ~\\ \hline
curve & $y^2=x^{2g+2}+t^2,~g~odd$ & $y^2=x^{2g+2}+t^2,~g~even$ \\ \hline
CB spectrum & $2,4,\ldots, g+1$ & $3,5\ldots, g+1$ \\ \hline
central charges& $a=\frac{1}{48}(g+2)(7g+5)$ & $a=\frac{1}{48} (g (7 g+19)+2)$ \\
&$c=\frac{1}{12}(g+2)(2g+1)$& $c=\frac{1}{12} (g (2 g+5)+1)$ \\ \hline
Singularities & $I_4I_1^{2g}$ & $I_4 I_1^{2g}$  \\ \hline
free hyper on CB & $g$& $g$ \\ \hline
flavor & $usp(2g)$ & $usp(2g)\times u(1)$ \\ \hline
\end{tabular}
 \begin{tabular}{|l|l|} \hline
  Full SW curve $g~odd$ &  $y^2=x^{2g+2}+(\tau x^{g+1}+u_{g-1} x^{g-1}+u_{g-3}x^{g-3}+\ldots+t)^2$ \\ ~& $+(T_2 x^{2g}+T_4x^{2g-2}+\ldots+T_{2g}x^2)$ \\ \hline
   Full SW curve $g~even$ &  $y^2=x^{2g+2}+(m x^{g}+u_{g-2} x^{g-2}+u_{g-4}x^{g-4}+\ldots+t)^2$ \\
   & $+(T_2 x^{2g}+T_4x^{2g-2}+\ldots+T_{2g}x^2)$\\ \hline
\end{tabular}
\end{center}
\label{twist1}
\end{table}%
 
 \begin{table}[H]
\caption{$\Z_2$ quotient: $y^2=t(x^{g+2}+t)$.}
\begin{center}
\begin{tabular}{|c|c|} \hline
Theory & $~$ \\ \hline
curve & $y^2=t(x^{g+2}+t),~g~even$   \\ \hline
CB spectrum & $4,6,\ldots, g+2$ \\ \hline
central charges& $a=\frac{1}{48} g (7 g+36)$  \\
&$c=\frac{1}{24} g (4 g+21)$ \\ \hline
Singularities & $I_4I_1^{2g+1}$  \\ \hline
free hyper & $\frac{7g}{2}-4$\\ \hline
flavor & $usp(2g+2) \times su(2)$  \\ \hline
  Full SW curve $g~even$ & $y^2=(T_2^\prime x^g+u_{g-2} x^{g-2}+\ldots+t) [x^{g+2}+(T_2^\prime x^g+u_{g-2} x^{g-2}+\ldots+t)]$ \\ 
  & $+[T_2 x^{2g+2}+T_4 x^{2g}+\ldots+T_{2g+2}x^2]$ \\ \hline
\end{tabular}
\end{center}
\label{twist2}
\end{table}%

\section{SW differential and other possibilities}
In our previous studies, we have assumed that the derivative of $\lambda_{sw}$ with respect to $t$ 
is given as ${\partial \lambda_{sw} \over \partial t}=c\frac{dx}{y}$. We would like to solve $\lambda_{sw}$ explicitly.
Let's assume the SW curve and SW differential take the following form
\begin{equation*}
y^2=f(x, P(x,u_i)),~~\lambda_{sw}=R(x,y, \sum_i P(x,u_i));
\end{equation*}
and the polynomial which contains the Coulomb branch operators could take the following general form
\begin{equation*}
P(x,u_i)=\sum_i u_i x^{i}.
\end{equation*}

Now let's consider the constraints on SW differential $\lambda_{sw}=R(P, x, y)\frac{dx}{y}$, and the derivative reads 
\begin{equation}
{\partial \lambda_{sw} \over \partial u_i}=[\frac{ \partial R}{\partial u_i} +\frac{R}{y^2}{\partial f \over \partial P} x^i] \frac{dx}{y}.
\end{equation}
We'd like to write it in the form of ${\partial \lambda_{sw} \over \partial u_i}=a {x^i dx \over y}+b d({x^{i+1}\over y})$. Using
\begin{equation}
d(\frac{x^{i+1}}{y})={(i+1)x^{i}dx\over y}-\frac{1}{y^2} x^{i+1} ({\partial f\over \partial P}P^{'}+{\partial f \over \partial x})\frac{dx}{y}.
\end{equation}
So if we'd like to derivative is given by a holomorphic differential, one has
\begin{equation}
\boxed{-x({\partial f \over \partial P} P^{'}+{\partial f \over \partial x})=R {\partial f \over \partial P}+c f.}
\label{crucial}
\end{equation}
Here $c$ is a constant. The final result would be
\begin{equation*}
{\partial \lambda_{sw} \over \partial u_i}= {\partial R \over \partial u_i}dx+[c-(i+1)] {x^i dx\over y} +d({x^{i+1} \over y}).
\end{equation*}
and the constant in equation \ref{crucial} is $c=-k$.

\textit{Example 1}: If $f$ is linear in $P$ (Type I and type II), the solution of SW differential is simple: $R=y^2$ an $\lambda_{sw}=ydx$. 
A total derivative term is not needed. Similarly for type III and type IV, one may need to take $R=\frac{y^2}{x}$, and $\lambda_{sw}=y \frac{dx}{x}$.

\textit{Example 2}: If $f$ takes the form $f=P^2+x^k$, the solution would be 
\begin{equation*}
R=\frac{k}{2}P-xP^{'}.
\end{equation*}

There exists an interesting possibility that involves the $D_n$ type conformal gauge theory. In this case, 
the polynomial encoding the Coulomb branch operator is written as the form
\begin{equation*}
P(x,u_i)=\sum u_i x^i+t^2.
\end{equation*}
and the derivative of $\lambda_{sw}$ with respect to $t$ is now proportional the "square root" of the differential $ \frac{dx}{y}$.

\textit{Example}: The curve is $y^2=f(x,t^2)=x^{4g}+(t^2)^2$, We impose a $\Z_2$ invariance $x\to -x$, and 
so the full SW curve is written as 
\begin{equation*}
y^2=x^{4g}+ (\tau x^{2g}+u_{2} x^{2g-2}+u_{2g-2} x^2+t^2)^2+ T_2x^{4g-2}+\ldots+T_{4g-4}x^4.
\end{equation*}
This gives the SW curve of $SO(2g)$ coupled with $2g-2$ fundamental hypermultiplets. Notice that the mass deformations from 
$x^2$ term and constant are not allowed \cite{Brandhuber:1995zp}. 
We might generalized above construction to the following curve ($g$ is odd)
\begin{equation*}
y^2=x^{2g}+(m x^{g-1}+\ldots+u_{g-2} x^2+t^2)^2+T_2x^{2g-2}+\ldots+T_{2g-4}x^4.
\end{equation*}
The Coulomb branch spectrum is $(3, 5, \ldots, g-2, \frac{g}{2})$, and the flavor symmetry is $usp(2g-4)\times U(1)$.
When $g=7$, a theory with same Coulomb branch spectrum and flavor symmetry was indeed found in \cite{Li:2022njl}.

\section{Conclusion}
We classify 4d $\mathcal{N}=2$ SCFTs whose SW geometry can be written as hyperelliptic family. 
The classification is made possible by assuming \ref{swcurve} and \ref{swdifferential}, and 
reduce the problem to one parameter family. It would be interesting to try to find other 
solutions to the special Kähler conditions.

Once the full SW family is given, one can further study the low energy physics of any vacuum \cite{Xie:2021hxd} by using the period integral.
Geometrically, one can also compute the flavor symmetry by computing the Mordell-Weil lattice \cite{shioda1999mordell}. 
One might also compute the torsion part of Mordell-Weil lattice which could be 
identified as higher form symmetry.

We focus on the SCFT in this paper, and the method can be applied to non conformal theory. 
The SW geometry can still be written as the form \ref{swcurve} and \ref{swdifferential}, and 
the only difference is that there is no $\mathbb{C}^*$ action acting on the one parameter family.

Another interesting generalization is to consider other algebraic curves such as order three 
algebraic curves $x^3=f(y,u_i)$. We hope to report the progress in this direction in the near future.


\bibliographystyle{JHEP}
\bibliography{ref}

\end{document}